\documentclass[prl,twocolumn,amsmath,showpacs]{revtex4}
\usepackage{graphicx}
\usepackage{color}

\begin{document}
\title{Point-contact study of $Re$FeAsO$_{1-x}$F$_{x}$ ($Re$=La, Sm) superconducting films}
\author{Yu. G. Naidyuk, O. E. Kvitnitskaya,  I. K. Yanson}
\affiliation{B. Verkin Institute for Low Temperature Physics and
Engineering, National Academy  of Sciences of Ukraine,  47 Lenin
Ave., 61103, Kharkiv, Ukraine}
\author{G. Fuchs, S. Haindl, M. Kidszun, L. Schultz, B. Holzapfel}
\affiliation{Leibniz-Institut f\"ur Festk\"orper- und
Werkstofforschung Dresden e.V., Postfach 270116, D-01171 Dresden,
Germany}
\date{\today}

\begin{abstract}
Point-contact (PC) Andreev-reflection (AR) measurements of the superconducting gap in iron-oxipnictide
$Re$FeAsO$_{1-x}$F$_{x}$ ($Re$=La, Sm) films have been carried out.
The value of the gap is distributed in the range $2\Delta \simeq$
5--10\,meV (for $Re$=Sm) with a maximum in the distribution around 6\,meV.
Temperature dependence of the gap $\Delta(T)$ can be fitted well
by BCS curve giving reduced gap ratio $2\Delta /kT_c^*\simeq 3.5$ (here $T_c^*$ is the critical temperature from the BCS fit).
At the same time, an expected second larger gap feature was difficult to resolve distinctly on
the AR spectra making determination reliability of the second gap detection questionable.
Possible reasons for this and the origin of other features like clear-cut asymmetry in the AR spectra
and current regime in PCs are discussed.



\end{abstract}

\maketitle
\section{Introduction}

Discovering a few years ago a new family of iron based superconductors gave rise both to intensive investigation of the fundamental properties of these materials (see reviews \cite{Sad,Ivan,Izy,Ishida}
and Refs. therein) and to seeking for their potential application.
The fabrication of high-quality thin films is of great importance for potential applications of these new materials in
superconducting devices as well as for a deeper fundamental study of their superconducting properties.
Numerous experiments were undertaken to study the superconducting state of iron based superconductors, however less
attention was paid to the investigation of films, due to the more complicated preparation.

In this paper we present first point-contact Andreev reflection (PCAR) spectroscopy investigation of
LaFeAsO$_{1-x}$F$_{x}$ and SmFeAsO$_{1-x}$F$_{x}$ films.
The main goal was to measure of the  superconducting gap(s) and its temperature dependence for the mentioned
films to compare these data with existing similar measurements on bulk samples \cite{Chen,Gonnelli,Wang}
and to clarify some issues as to the presence of multiband structure on PCAR spectra and PCAR spectra asymmetry.

\section{Experimental details}
\begin{figure} [b]
\vspace{3cm}
\begin{center}
\includegraphics[width=8.5cm,angle=0]{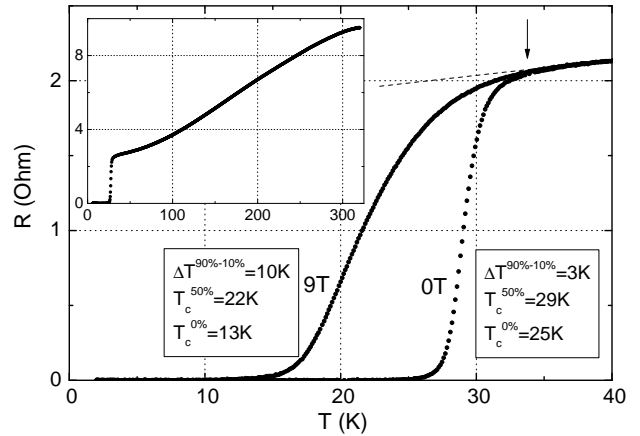}
\vspace{-3cm}
\end{center}
\vspace{0cm} \caption[] {Superconducting transition
of a SmFeAsO$_{1-x}$F$_{x}$ thin film at zero field and at 9\,T. Arrow shows $T_c^{onset}\simeq 34\,K$.
Inset: Temperature dependence of the resistance of the film up to room temperatures.} \label{rt}
\end{figure}

$Re$FeAsO$_{1-x}$F$_{x}$ ($Re$=La, Sm) films with superconductivity onset below 34\,K  have
been fabricated using pulsed laser deposition. The successful growth of high quality $Re$FeAsO$_{1-x}$F$_{x}$
thin films \cite{Backen,Kidszun,Kidszun1,Kidszun2} enables the investigation of fundamental properties.
Details of the preparation of LaFeAsO$_{1-x}$F$_{x}$   films are described in Ref. \cite{Kidszun}.
The same parameters were also used for SmFeAsO$_{1-x}$F$_{x}$ film growth.
The temperature dependence of the resistance of the $Re$=Sm film used in our experiments is
shown in Fig.\,\ref{rt}.
The onset of the superconducting transition of about 34\,K is well below the maximal transition temperature
for optimally doped SmFeAsO$_{1-x}$F$_{x}$ crystals likely due to severe fluorine loss under film preparation.
Also the superconducting transition sufficiently broadens in a magnetic field, what can be due to nonuniform
fluorine doping through the film thickness. The $Re$=La film was of lower quality with a much broader
superconducting transition. Therefore, we concentrated  in this study mainly on measurements of  the $Re$=Sm film.

The PCs were established by the standard "needle-anvil" method \cite{Naid} touching of
the film surface by a sharpened thin Cu wire. The differential resistance $dV/dI(V)$
was recorded by sweeping the dc current $I$ on which a small ac current $i$ was superimposed
using the standard lock-in technique. The measurements of PCAR spectra were performed in the temperature range
between 3 and 33\,K applying a magnetic field up to 7\,T perpendicular to the film surface.

\section{Characteristic lengths}

In PCAR spectroscopy some conditions must be fulfilled, namely, the contact diameter $d$ has to be smaller
than the inelastic electron mean-free path $l_i$ as well as smaller than the superconducting coherence length $\xi$.
The latter is anisotropic in the $Re$FeAsO$_{1-x}$F$_{x}$ family and does not exceed a few nm \cite{Ishida}.
The contact size can be estimated from its resistance $R_{\rm {PC}}$ above $T_c$ using modified Wexler formula \cite{Naid}
for the case of heterocontact with a metal, where its resistivity is negligible compared with that of $Re$FeAsO$_{1-x}$F$_{x}$:
\begin{equation}
R_{\rm {PC}}\simeq (1+Z^2)\frac{16\rho l}{3 \pi d^2}+ \frac{\rho}{2d}.
 \label{Wexler}
\end{equation}
Here, $Z$ is the so called barrier strength \cite{BTK}. We estimated the product $\rho l$ using the upper limit
for the electron carrier concentration $n_e\leq 10^{21} $cm$^{-3}$ as inferred from Hall data
\footnote{The Hall coefficient determined in \cite{Jaros} for $Re$=La is about 3 times lager
at low temperatures than that for $Re$=Sm, which points to a 3 times larger electron density in $Re$=Sm.
At the same time for $Re$=Sm with an onset of superconductivity at 36\,K, similar as
in our film, the electron density is estimated to be about $n_e\leq 0.6\cdot 10^{21} $cm$^{-3}$ \cite{Tropeano}.}
for $Re$=La just above $T_c$ \cite{Sefat}. Making use of the Drude free electron model we calculated
$\rho l \simeq 1.3\cdot10^{4} n_e^{-2/3} = 1.3\cdot 10^{-10}\Omega\cdot$cm$^{2}$.
The residual resistivity $\rho_0$ of $Re$FeAsO$_{1-x}$F$_{x}$  compounds, estimated by extrapolation $\rho(T\to0)$, has value of about 100\,$\mu\Omega\cdot$cm for $Re$=La, whereas $\rho_0$ is higher for other $Re$FeAsO$_{1-x}$F$_{x}$  compounds
\footnote{According to \cite{Jaros}, the resistivity of $Re$=Sm is 2-3 times larger at low temperatures
than that of $Re$=La.}. Using the calculated value of $\rho l$ and the mentioned resistivity $\rho_0$, we find 10\,nm
as upper limit for the elastic electron mean free path $l_e$ for $Re$FeAsO$_{1-x}$F$_{x}$ compounds.

In Table I, we present estimations of the PC diameter, according to Eq.\,(\ref{Wexler}), for PCs with three different
resistances using both the lower limit of $\rho_0$=100 $\mu\Omega\cdot$cm (for high quality samples) and the higher
value $\rho_0$=1\,m$\Omega\cdot$cm, taken from the literature for $Re$FeAsO$_{1-x}$F$_{x}$.
\begin{table}[b]
\caption{Calculation of the PC diameter for three PC resistances according to Eq.\,(\ref{Wexler}) for $Z$=0 and two values
of resistivity and for $Z$=0.6 (last column).}
\begin{tabular}{|c|c|c|c|}
\hline
  $R_{\rm {PC}}$ & $\rho_0$=100 $\mu\Omega\cdot$cm & $\rho_0$=1 m$\Omega\cdot$cm & $\rho_0$=100 $\mu\Omega\cdot$cm, $Z$=0.6 \\
\hline
  10$\Omega$ & 78\,nm & 504\,nm &85\,nm \\
  30$\Omega$ & 37\,nm & 171\,nm &41\,nm \\
  100$\Omega$& 18\,nm & 54\,nm &20\,nm  \\
\hline
\end{tabular}
\end {table}
Thus, in order to fulfil the condition that the dimension of the PC has to be smaller than the elastic mean free path
of electrons and/or the coherence length the PC resistance must be well above a few hundred Ohms,
which is not the case for PCs both investigated in this paper and in existing
PCAR studies cited in Ref.\,\cite{Gonnelli}. Therefore all PCAR measurements have
been done at least in the diffusive regime when elastic mean free path is small, i. e. $d\gg l_e $. Although the diffusive regime
does not prevent PCAR spectroscopy, it favors the shortening of the diffusive inelastic
mean free path $\Lambda\simeq (l_e\cdot l_i)^{1/2}$ of electrons so that a transition to the thermal regime $d\gg $min$(l_i, \Lambda$) \cite{Naid,Verkin79} with increasing
of applied voltage is getting more probable. Taking into account that the resistivity in all $Re$FeAsO$_{1-x}$F$_{x}$  has a remarkable
slope (increase) just above $T_c$, e. g., for $Re$=La,  $\rho(T)$ behaves like $T^2$ below $\sim$ 200\,K \cite{Sefat},
the inelastic, likely electron-electron, scattering in $Re$FeAsO$_{1-x}$F$_{x}$  is starting already at very low temperatures shortening an inelastic mean free path $ l_i$ and $\Lambda$.
\begin{figure} [t]
\begin{center}
\includegraphics[width=8cm,angle=0]{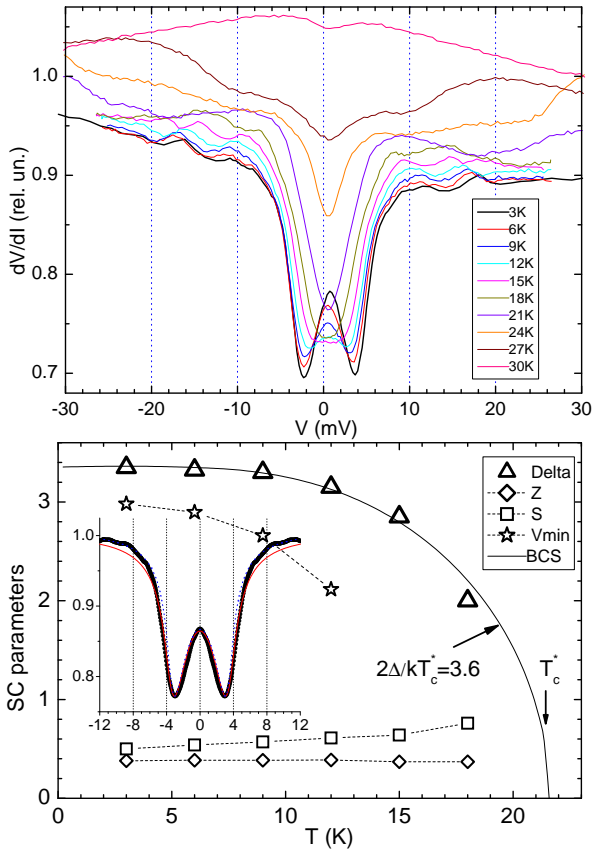}
\end{center}
\vspace{0cm} \caption[] {(Color on-line) Upper panel: $dV/dI$ curves of a
SmFeAsO$_{1-x}$F$_{x}$ - Cu contact ($R$ = 26\,$\Omega$)
for varying temperature.  Bottom  panel: Temperature dependencies of the fitting
parameters: superconducting gap $\Delta$ (triangles), barrier parameter $Z$ (diamonds),
scaling parameter $S$ (squares), position of minima in $dV/dI$ (stars) for the PC
from the upper panel. The broadening parameter $\Gamma$ is equal
to zero. The solid line represents the BCS-like gap behavior.
Inset: symmetrized and normalized \footnote{Resistivity of Sm and La pnictides has pronounced temperature dependence even at low temperatures and also below $T_c$, as it follows from the measurement in a magnetic field, see, e.g., Ref.\,\cite{Kidszun2}. It means that, so-called, normal state curve used for normalization can be slightly different for different temperatures. Therefore we have used for normalization of each $dV/dI(V)$ a weak parabolic-like curve which fitted the same $dV/dI(V)$ at biases $|V|>$10\,mV, i.e. above the gap minima.} $dV/dI$ curve at
$T$ = 3\,K (points) together with curves calculated according to
the generalized BTK theory: dashed (blue) and solid (red) lines - fit with $\Delta$=3.35\,mV, $\Gamma$=0, $Z$=0.38, $S$=0.5 and
$\Delta$=3.47\,mV, $\Gamma$=0.25\,mV, $Z$=0.41, $S$=0.58, respectively.} \label{dvdi}
\end{figure}

On the other hand, the coherence length $\xi$ is also much smaller than the contact size,
what can result in a distribution of superconducting properties (e.\,g., critical temperature, gap value) within
the PC and in a suppression of superconductivity in some part of the PC with current increase.
Thereby, regardless of the PC resistance (of course the higher the resistance the higher
probability to be in the spectral regime), each $dV/dI(V)$ curve should be
critically analyzed if it suits for spectroscopy of the superconducting gap.

\section{PCAR spectroscopy of the superconducting energy gap in S{\tiny m}F{\tiny e}A{\tiny s}O$_{1-x}$F$_{x}$}

Because of mentioned above reasons, it was difficult to get "clean" PCAR $dV/dI$ spectra having no
humps, spikes and other irregularities, which are connected with deviations from the spectral
regime in PC. Fig.\,\ref{dvdi} shows one of the best series of  $dV/dI$ curves,
which demonstrate clear Andreev-reflection (gap) structure, namely pronounced minima at $V\simeq \pm3$\,mV
at $T\ll T_{\rm c}$ which are to a great extend free from unwanted features.
To get the superconducting gap value $\Delta$ and other parameters from the PCAR spectra
the Blonder-Tinkham-Klapwijk (BTK) theory \cite{BTK} including so-called
broadening parameter $\Gamma$ has been used.
The temperature dependence of the superconducting gap $\Delta$ has been established (see Fig.\,\ref{dvdi})
from the fitting within the BTK theory. We tried to keep constant such fitting parameters as $\Gamma$,
the barrier $Z$ and the scaling $S$.
\begin{figure} [t]
\begin{center}
\includegraphics[width=8cm,angle=0]{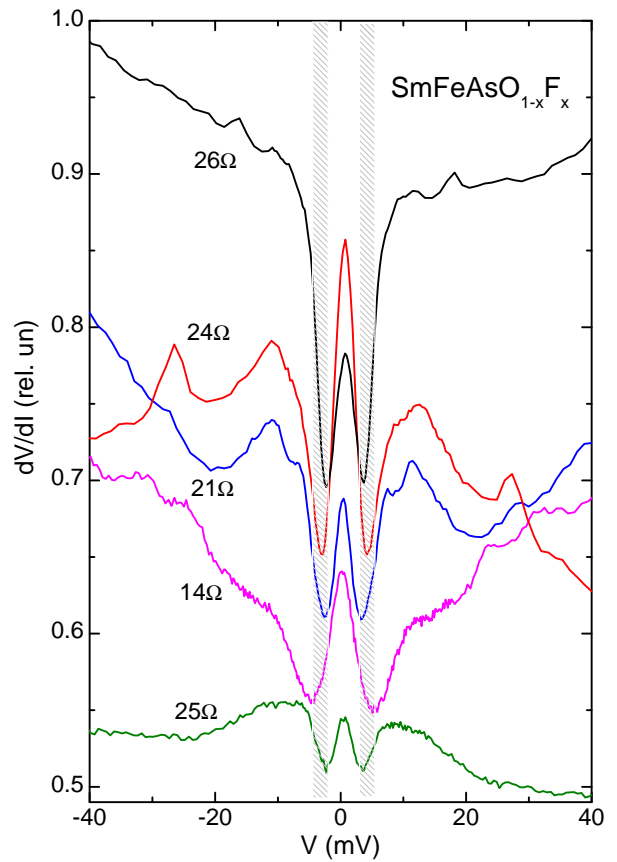}
\end{center}
\vspace{0cm} \caption[] {$dV/dI$ curves of several SmFeAsO$_{1-x}$F$_{x}$ -- Cu contacts measured at 3\,K.
PC resistance is shown for each curve. Vertical dashed stripes mark reproducible AR minima.} \label{dvdis}
\end{figure}
Here, the parameter $S$ corresponds to the ratio of the experimental $dV/dI$ intensity to the calculated one
\footnote{Parameter S is intended to equalize the intensities of experimental and theoretical curves.
Theoretically $S$ = 1, but $S <$ 1 happens very often too. The possible reasons are connected with deviation of the PC
structure from theoretical model and are described in more detail in Appendix of Ref.\,\cite{Bobrov}.}.

It is seen from  Fig.\,\ref{dvdi} that $\Delta(T)$ is in line with the BCS curve (solid line),
however, the resulting critical temperature of $T_{\rm c}^{*}\simeq$ 21.6\,K obtained by extrapolating the BCS curve
to $\Delta$= 0 is significantly lower than $T_{\rm c}$ determined from the temperature dependence of the resistivity
of the SmFeAsO$_{1-x}$F$_{x}$ film (see Fig.\,\ref{rt}). At the same time, the superconducting main minimum in
$dV/dI$ for the investigated PC disappears close to 30\,K as shown in Fig.\,\ref{dvdi} which approximately
corresponds to the midpoint ($T_{\rm c}^{50\%}$) of the superconducting transition in Fig.\,\ref{rt}. However, by approaching
this temperature, the shape of the $dV/dI$ minimum starts to deviate from the theoretically expected behavior
already above 20\,K as shown in Fig.\,\ref{dvdi}.
This is likely due to inhomogeneities of the superconducting state in the PC region in the case if its size $d$ is larger
than the coherence length $\xi$. Also with increasing temperature inelastic scattering increases favoring the
transition to the thermal regime. This hinders a determination of the gap in this temperature range. Therefore, the
determination of the gap of the PC shown in Fig.\,\ref{dvdi} is restricted to temperatures below 20\,K.

\begin{figure} [t]
\vspace{3cm}
\begin{center}
\includegraphics[width=8cm,angle=0]{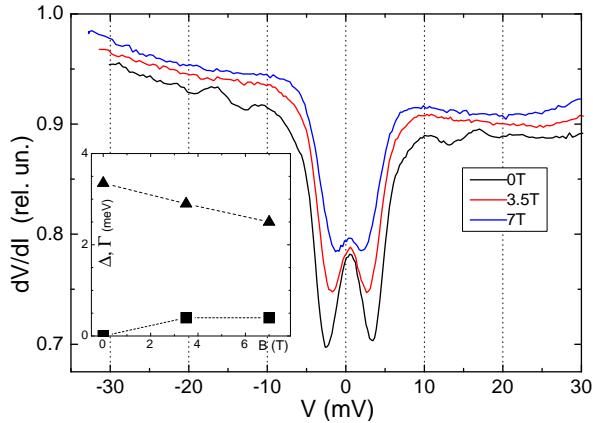}
\vspace{-3cm}
\end{center}
\vspace{0cm} \caption[] {$dV/dI$ curves of PC from Fig.\,\ref{dvdi} measured at 3\,K in a magnetic field.
Inset shows dependence of $\Delta$ (triangles) and $\Gamma$ (squares) vs magnetic field as determined from the
BTK-fitting of $dV/dI$ curves from the main panel. Here, $Z\simeq$0.38 and $S\simeq$0.5. Linear extrapolation of
$\Delta$  to zero result in a critical field of about 30\,T.} \label{dvdim}
\end{figure}

Investigating a dozen of PCs with clear double minima AR structure around $\pm 3$\,mV (see, Fig.\,\ref{dvdis}), we,
nevertheless, could not observe or detect reproducible structures at a higher voltage, which were found, e.g., in the representative two-band
superconductor MgB$_2$ \cite{Naid1}. Also an about 3 times larger second gap features reported for bulk SmFeAsO$_{1-x}$F$_{x}$
samples \cite{Gonnelli} could not be unambiguously resolved for the investigated PCs of the SmFeAsO$_{1-x}$F$_{x}$ film. Instead,
for increasing bias usually a peaked structure or other irregularities appear in the $dV/dI$ characteristics testifying
the transition to the non-spectral regime. But even if the similar to the second gap features are observed as, e.g,
shallow minima slightly above $\pm 10$\,mV in the upper PC spectrum in Fig.\,\ref{dvdis} (see also low temperature spectra in Fig.\,\ref{dvdi}), then this gap structure expected around $\pm$(9-12)\,meV is hardly to resolve for the other PC spectra (see, e.g., the bottom spectrum in Fig.\,\ref{dvdis}).
Moreover, already a weak magnetic field suppresses the mentioned structure, nevertheless no presence of any second
gap features appears (see Fig.\,\ref{dvdim}). The same concerns $dV/dI$ of PC with $R =14\,\Omega$, where
a shoulder is seen above $\pm$ 10\,mV often taken as a larger gap structure. The shoulder is washed out in magnetic
field 7\,T, while the small-gap minima are only slightly suppressed at the same field. On the contrary, in the mentioned MgB$_2$
the small gap is vanished more quickly in magnetic field than the larger one (see circles in Fig.\,15 in \cite{YansonR}).


The above mentioned larger gap structure is also badly resolved in a recent tunneling study of  SmFeAsO$_{1-x}$F$_{x}$ compounds
\cite{Noat}. The authors concluded that interband quasiparticle scattering has a crucial effect on the shape of the
tunneling spectra smearing out in particular a larger gap structure in the electronic DOS. Taking into account the very short
elastic electron mean free path in SmFeAsO$_{1-x}$F$_{x}$ compounds the interband quasiparticle scattering is expected to be quite strong.

From another tunneling measurements \cite{Fasano}, a distribution of gaps (conductance peaks) $\Delta_p$  between 6 and 8\,meV
with a mean value of 7\,meV was derived, which results in a reduced gap 2$\Delta_p /kT_c\simeq3.6$. Also from these tunneling
spectra, no second gap structure was resolved.

\begin{figure} [t]
\vspace{3cm}
\begin{center}
\includegraphics[width=8cm,angle=0]{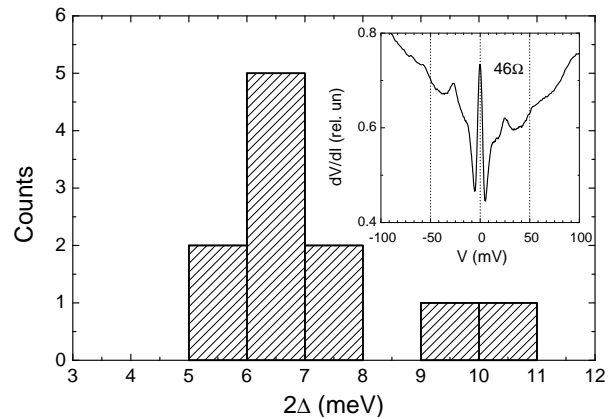}
\vspace{-3cm}
\end{center}
\vspace{0cm} \caption[] {Histogram of the gap distribution determined from $dV/dI(V)$ of SmFeAsO$_{1-x}$F$_{x}$ -- Cu contacts
with pronounced double minimum AR structure. Inset shows an example of $dV/dI(V)$ for PC with $2\Delta\simeq$10\,meV.}
\label{his}
\end{figure}

We show in Fig.\,\ref{his} a histogram of the gap distribution measured for PCs with the pronounced double minimum AR structure.
The distribution has a maximum around $2\Delta$=6\,mV and a high energy tail around 10\,mV. We connect this
large gap value and its broad distribution with the inhomogeneities of the superconducting properties on the film
surface due to fluorine loss. Supposing that the critical  temperature for the PC with the gap of
around 10\,meV is close to the onset of superconductivity of $\simeq$ 34\,K in  Fig.\,\ref{rt},
then $2\Delta /kT_c$ will be again close to the BCS value of 3.5.

\section{Asymmetry of $dV/dI$ curves}
\begin{figure} [t]
\begin{center}
\includegraphics[width=8cm,angle=0]{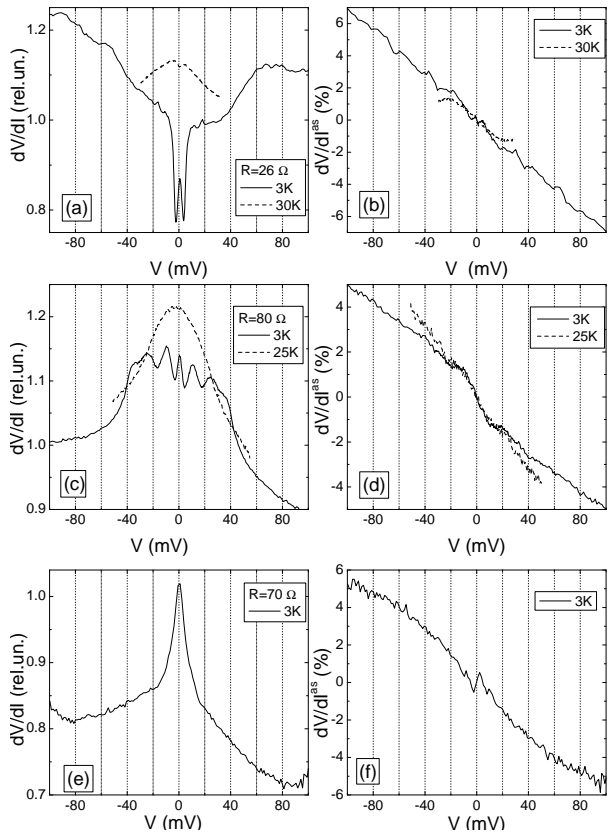}
\end{center}
\vspace{0cm} \caption[] {Left panels: $dV/dI$ curves of a SmFeAsO$_{1-x}$F$_{x}$ -- Cu contacts measured at 3\,K (solid curves)
for a broader bias range. Dashed curves have been measured for the same contacts at higher temperature.  PC in panel (a) is
the same as shown in Fig.\,\ref{dvdi}. Right  panels: antisymmetric part  $dV/dI^{as}(\%)= 100R_{PC}^{-1}(dV/dI(V>0)-dV/dI(V<0))$
for the corresponding curves from the left panels. Voltage polarity is given for the film.} \label{as}
\end{figure}
A characteristic feature of $dV/dI(V)$ of PCs with $Re$FeAsO$_{1-x}$F$_{x}$ is their pronounced asymmetry. From our study of this phenomenon
we conclude that the asymmetry of $dV/dI(V)$ is present independent of the superconducting features. Fig.\,\ref{as}
shows $dV/dI(V)$ curves for a few PCs, where clear AR double minima are observed (Fig.\,\ref{as}a) or the  superconducting
structure is superimposed on a broad maximum (Fig.\,\ref{as}c) or no superconducting features, but only a sharp zero-bias
maximum is present (Fig.\,\ref{as}e).
In all cases, the antisymmetric  part of $dV/dI(V)$ shows similar behavior and even the same relative value independent
of the drastic difference in $dV/dI(V)$ shape corresponding to different mechanisms of $dV/dI(V)$ formation. Such robustness
of the asymmetry of $dV/dI(V)$ testifies that this phenomenon is not related to the PC properties, but mainly determined by
the properties of the bulk material (or the film). One common and plausible reason of $dV/dI(V)$ asymmetry can be the thermoelectric
(Seebeck) effect if the PC is heated up at increasing bias in the thermal regime \cite{Naid}. It is known that $Re$FeAsO$_{1-x}$F$_{x}$ compounds
have a huge Seebeck coefficient at low temperatures which reaches -80\,$\mu$V$\cdot$K$^{-1}$ for $Re$=Sm just above
$T_c$ for a sample with $T_c$=53\,K \cite{Matus} and even more (-140\,$\mu$V$\cdot$K$^{-1}$) for a sample with $T_c$=36\,K \cite{Tropeano}
(similar as in our film). Taking thermoelectric effects in the case of heterocontacts  into account, the $dV/dI(V)$ asymmetry
was explained both for heavy-fermion high-resistive compounds \cite{Naid,Naid2} and for diluted Kondo-alloys \cite{Naid,Naid3},
where the Seebeck coefficient is strongly enhanced.
It seems that an equilibrium heating of the PC is not the only condition for the appearance of thermoelectric effects. The electron
distribution function in ballistic and diffusive PCs is in a highly nonequilibrium state which can be represented by two Fermi
spheres shifted from each other by applied voltage e$V$  \cite{Naid}. This nonequilibrium electronic state produces nonequilibrium
phonons with an effective temperature $\leq$eV/4 \cite{Kulik} which is close to the temperature eV/3.63 \cite{Verkin79} reached in
the thermal regime.
Therefore, we suggest that the nonequilibrium state in the PC can also cause a thermoelectric voltage in the case of heterocontacts.
This results in a $dV/dI(V)$ asymmetry similar as observed in PCs in the thermal regime \cite{Naid}.

We also would like to draw attention that asymmetry of the STM spectra, see Figs.\,5, 6 in Ref.\,\cite{Noat}, variates and even has different sign for the spectra with similar structure. Since STM spectra reflects in a direct way electronic density of states (DOS), then the non-reproducible asymmetry of STM spectra point out also to the non-DOS nature of the asymmetry, what strengthen our
arguments.

\section{PCAR spectroscopy of the superconducting energy gap in L{\tiny a}F{\tiny e}A{\tiny s}O$_{1-x}$F$_{x}$}

\begin{figure} [t]
\vspace{3cm}
\begin{center}
\includegraphics[width=8cm,angle=0]{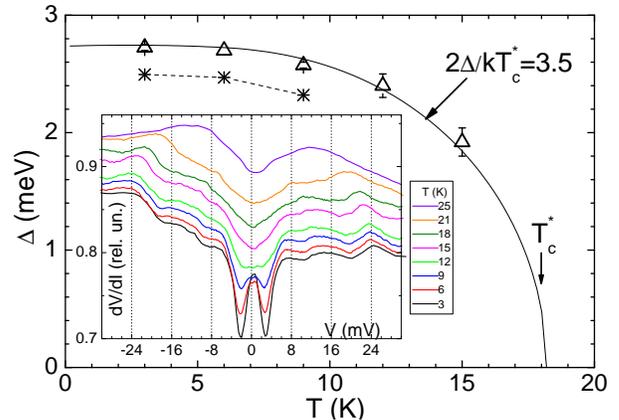}
\vspace{-3cm}
\end{center}
\vspace{0cm} \caption[] {(Color on-line) Temperature dependence of the superconducting gap $\Delta$ (triangles)
from the fitting of $dV/dI$ curves shown in  the inset. The position of minima in $dV/dI$ is marked by stars.
The solid line represents the BCS-like gap behavior. Inset: $dV/dI$ curves of a
LaFeAsO$_{1-x}$F$_{x}$ -- Cu contact ($R$ = 10\,$\Omega$)
for varying temperature. } \label{la}
\end{figure}

As mentioned above, the investigated LaFeAsO$_{1-x}$F$_{x}$  film has a rather poor quality showing a broad superconducting transition already at zero-field which is comparable with that of the SmFeAsO$_{1-x}$F$_{x}$ film at 9\,T (see Fig.\,\ref{rt}). Additionally, a kink appears in $\rho(T)$) at temperatures around 18\,K. The inset in Fig.\,\ref{la} shows $dV/dI$ of LaFeAsO$_{1-x}$F$_{x}$ -- Cu PC with Andreev-reflection minima for varying temperature along with gap behavior evaluated by fitting of the $dV/dI$ curves. Similar as for the SmFeAsO$_{1-x}$F$_{x}$ film
only small gap minima are found.
Again, the critical temperature obtained by extrapolating the BCS curve to $\Delta$=0 is low, i. e. about 18\,K (close to the kink in $\rho(T)$), while superconducting features in $dV/dI$ are seen even at 25\,K as shown in Fig.\,\ref{la}. We attribute this behavior with
the variation of the superconducting properties in the PC region due to the short coherence length and the sample
inhomogeneity.

\section{Conclusion}
We investigated the superconducting energy gap in $Re$FeAsO$_{1-x}$F$_{x}$ ($Re$=La, Sm) films by PCAR spectroscopy.
The mean value of the reduced superconducting gap 2$\Delta/kT_c^*$ is found to be close to the BCS value of 3.5
what is in agreement with PCAR data reported in \cite{Chen}. A second larger gap feature is hardly to resolve on PCAR spectra.
The reason can be strong interband quasiparticle scattering, which smears out the larger gap structure in the electronic
DoS both in the tunneling \cite{Noat,Fasano} and in the PCAR spectra. Note also that all superconducting features in $dV/dI$,
including these of non-Andreev nature, have disappeared above $T_c^{onset}\simeq 34\,K$. 

We demonstrate that the asymmetry of $dV/dI(V)$ curves is independent of the shape of the curves or whether superconducting
or AR features are present or not in $dV/dI(V)$. Therefore, the asymmetry is not related to the spectroscopy but
reflects bulk properties of $Re$FeAsO$_{1-x}$F$_{x}$. Very probably the asymmetry is caused by the thermoelectric effect.

\section*{Acknowledgements}
Two of the authors (Yu.G.N. $\&$ O.E.K.) would like to thank the Alexander von Humboldt Foundation for the support
and Dr. K. Nenkov for the technical assistance. S.H. and M.K. would like to acknowledge the financial support by DFG
(under project HA5934/1-1).



\begin{thebibliography}{99}

\bibitem{Sad} Sadovskii M V 2008 Physics-Uspekhi {\bf 51} 1201

\bibitem{Ivan} Ivanonskii A L 2008 Physics-Uspekhi {\bf 51} 1229

\bibitem{Izy} Izyumov Y A and Kurmaev E Z 2008 Physics-Uspekhi {\bf 51} 1261

\bibitem{Ishida}
Ishida K, Nakai Y and Hideo Hosono 2009 J. Phys. Soc. Japan {\bf 78} 062001

\bibitem{Chen}
Chen T Y, Tesanovic Z, Liu R H, Chen X H and Chien C L 2008 Nature {\bf 453} 1224;
Chen T Y, Huang S X, Tesanovic Z, Liu R H, Chen X H and Chien C L 2009
Physica C {\bf 469} 521

\bibitem{Gonnelli} Gonnelli R S, Daghero D, Tortello M, Ummarino G A, Stepanov V A, Kremer R K, Kim J S,
Zhigadlo N D, Karpinski J 2009 Physica C {\bf 469}  512;  Daghero D, Tortello M,  Gonnelli R S,
Stepanov V A, Zhigadlo N D and Karpinski J 2009 Phys. Rev. B {\bf 80} 060502(R)

\bibitem{Wang} Wang Y-L, Shan L, Fang L, Cheng P, Ren C and Wen H-H 2009
Supercond. Sci. Technol. {\bf 22} 015018

\bibitem{Backen} Backen E, Haindl S, Niemeier T, H\"uhne R, Freudenberg T,
Werner J, Behr G, Schultz L and Holzapfel B 2008 Supercond. Sci. Technol. {\bf 21} 122001

\bibitem{Kidszun} Kidszun M, Haindl S, Thersleff T, Werner J, Langer M, H\"anisch J, Iida K, Reich E,
Schultz L and Holzapfel B 2010 Europhys. Lett. {\bf 90} 57005

\bibitem{Kidszun1} Kidszun M, Haindl S, Reich E, H\"anisch J, Iida K, Schultz L and
Holzapfel 2010 Supercond. Sci. Technol. {\bf 23} 022002

\bibitem{Kidszun2} Kidszun M, Haindl S, Thersleff T, H\"anisch J, Kauffmann A, Iida K,
Freudenberger J, Schultz L, Holzapfel B ($Preprint$ arXiv:1004.4185)

\bibitem{Naid}
Naidyuk Yu G and Yanson I K 2004  {\it Point Contact Spectroscopy},
Springer Series in Solid-State Sciences, Vol.145 (Springer, New York)


\bibitem{BTK}
Blonder G E, Tinkham M and Klapwijk T M 1982 Phys. Rev. B {\bf 25} 4515

\bibitem{Bobrov} Bobrov N L, Chernobay V N, Naidyuk Yu G, Tyutrina L V, Yanson I K, Naugle D G,
Rathnayaka K D D 2010 Sov. J. Low Temp. Phys. {\bf 36} 990 [2010 Fiz. Nizk. Temp. {\bf 36} 1228]

\bibitem{Jaros} Jaroszynski J, Riggs S C, Hunte F, Gurevich A, Larbalestier D C, Boebinger G S,
Balakirev F F, Migliori A, Ren Z A, Lu W, Yang J, Shen X L, Dong X L, Zhao Z X, Jin R, Sefat A S,
McGuire M A, Sales B C, Christen D K, and Mandrus D 2008 Phys. Rev. B {\bf 78} 064511

\bibitem{Verkin79}
Verkin B I, Yanson I K, Kulik I O, Shklyarevski O I, Lysykh A A
and Naidyuk Yu G 1979 Sol. State Commun. {\bf 30} 215

\bibitem{Sefat}
Sefat A S, McGuire M A, Sales B C, Jin R, Howe J Y and Mandrus D
2008 Phys. Rev. B {\bf 77} 174503

\bibitem{Naid1}
Naidyuk Yu G, Yanson I K, Tyutrina L V, Bobrov N L, Chubov P N,
Kang W N, Kim H-J, Choi E-M, and Lee S-I 2002 Sov. Phys. -- JETP Lett. {\bf 75} 238

\bibitem{YansonR} Yanson I K, Naidyuk Yu G 2004 Low Temp. Phys. {\bf 30} 261
[2004 Fiz. Nizk. Temp. {\bf 30}, 355]

\bibitem{Noat} Noat Y, Cren T, Dubost V,  Lange S, Debontridder F, Roditchev D, Marcus J and Toulemonde P,
Sacks W 2010 J. Phys.: Condens. Matter {\bf 22} 465701

\bibitem{Fasano} Fasano Y, Maggio-Aprile I, Zhigadlo N D, Katrych S, KarpinskiJ  and Fischer {\O} 2010
Phys. Rev. Lett., {\bf 105} 167005

\bibitem{Matus} Matusiak M, Plackowski T, Bukowski Z, Zhigadlo N D and Karpinski J
2009 Phys. Rev. B {\bf 79} 212502

\bibitem{Naid2} Naidyuk Yu G and Yanson I K 1998 J.~Phys.: Condens.~Matter {\bf 10} 8905

\bibitem{Naid3} Naidyuk Yu G, Gribov N N,  Shklyarevskii O I, Jansen A G M  and
Yanson I K 1985 Sov. J. Low Temp. Phys. {\bf 11} 580 [1985 Fiz. Nizk. Temp. {\bf 11} 1053]

\bibitem{Kulik}
Kulik I O, Yanson I K, Omelyanchouk A N 1981 Sov. J. Low Temp. Phys. {\bf 7} 129
[1981 Fiz. Nizk. Temp. {\bf 7} 263]

\bibitem{Tropeano} Tropeano M, Fanciulli C, Ferdeghini C, Marr`e D, Siri A S,
Putti M, Martinelli A, Ferretti M, Palenzona A,  Cimberle M R, Mirri C, Lupi S, Sopracase R,
Calvani P and Perucchi A 2009 Supercond. Sci. Technol. {\bf 22} 034004

\end{thebibliography}
\end{document}